\documentclass[preprint,12pt]{elsarticle}

\usepackage{comment}
\usepackage{amssymb}
\usepackage{amsmath}
\usepackage{xurl}
\usepackage{float}
\usepackage{hyperref}
\hypersetup{
    colorlinks=true,
    anchorcolor=black,
    citecolor=blue,
    linkcolor=blue,
    filecolor=blue,      
    urlcolor=blue,
}

\journal{Preprint}

\begin{document}

\begin{frontmatter}



\title{The MUG-10 Framework for Preventing Usability Issues in Mobile Application Development}


\author[inst1]{Pawel Weichbroth}
\affiliation[inst1]{organization={Department of Software Engineering, Faculty of Electronics, Telecommunications and Informatics, Gdansk University of Technology},
            addressline={Narutowicza 11/12}, 
            city={Gdansk},
            postcode={80-233}, 
            state={pomorskie},
            country={Poland} \\
            email: pawel.weichbroth@pg.edu.pl}

\author[inst2]{Tomasz Szot}
\affiliation[inst2]{organization={Gdansk University of Physical Education and Sport},
            addressline={Kazimierza Górskiego 1}, 
            city={Gdansk},
            postcode={80-336}, 
            state={pomorskie},
            country={Poland}}

\begin{abstract}
Nowadays, mobile applications are essential tools for everyday life, providing users with anytime, anywhere access to up-to-date information, communication, and entertainment. Needless to say, hardware limitations and the diverse needs of different user groups pose a number of design and development challenges. According to recent studies, usability is one of the most revealing among many others. However, few have made the direct effort to provide and discuss what countermeasures can be applied to avoid usability issues in mobile application development. Through a survey of 20 mobile software design and development practitioners, this study aims to fill this research gap. Given the qualitative nature of the data collected, and with the goal of capturing and preserving the intrinsic meanings embedded in the experts' statements, we adopted in vivo coding. The analysis of the collected material enabled us to develop a novel framework consisting of ten guidelines and three activities with general applications.
In addition, it can be noted that active collaboration with users in testing and collecting feedback was often emphasized at each stage of mobile application development. Future research should consider focused action research that evaluates the effectiveness of our recommendations and validates them across different stakeholder groups. In this regard, the development of automated tools to support early detection and mitigation of usability issues during mobile application development could also be considered.
\end{abstract}




\end{frontmatter}

\section{Introduction}
\label{sec:introduction}
In the realm of ubiquitous wireless networking and mobile devices, mobile application development is constantly evolving to meet the growing requirements of users \cite{Statista2025}. On the other hand, the mobile app market is considered both attractive and profitable \cite{Wagner2023}, with a global valuation of USD 252.89 billion in 2023 with over 6.3 billion users across the world \cite{GVR}. While mobile applications have penetrated almost every area of human activity, from health to education \cite{gawlik2023education}, finance \cite{garcia2024explainable}, communication \cite{wang2022vision}, and entertainment \cite{falkowski2020current}, one of the main areas of ongoing research concerns usability \cite{ali2022mobile,salman2018usability}. 

In the realm of mobile applications, usability is typically understood in terms of the ISO 9241-11 standard \cite{weichbroth2020usability}, which defines usability as the extent to which a product (system, software, or service) can be used by specified users to achieve their goals with effectiveness, efficiency, and satisfaction in a specified context of use \cite{ISO1998}. However, over the years, usability has also been evaluated based on various other attributes that reflect the specific needs and requirements of different user groups \cite{daif2018mobile,weichbroth2018usability,acosta2021accessibility,iakovets2022use}. Although a customized app can better align with users' needs and objectives, it can also introduce new usability challenges. 

Undeniably, both the academic and software development communities are still developing and testing different methods and tools to address a variety of usability issues in an effective way \cite{huang2023systematic,kumar2024usability}. Although recent studies report the results of usability testing across various applications \cite{akmal2021development,storm2021usability}, few have addressed actionable and preventive countermeasures \cite{weichbroth2024usability}. The present study attempts to fill this research gap. More specifically, our goal is to identify recent advances in this area through interviews with a group of experts. To this end, we aim to develop and introduce a generic, actionable framework that offers a new perspective on the matter.

The rest of the papers organized as follows. 
Section~\ref{sec:background} provides the theoretical background.
Section~\ref{sec:methodology} outlines the research methodology.
Section~\ref{sec:results} presents the MUG-10 framework.
Section~\ref{sec:framework-eva} evaluates the introduced MUG-10 framework.
Section~\ref{sec:discussion} discusses the limitations and contributions of the study.
Section~\ref{sec:conclusion} concludes the paper and suggests future research directions.

\section{Background}
\label{sec:background}
By its very nature, usability as such does not exist \cite{weichbroth2024usability}. It arises as a result from the user's conscious experience with a mobile application \cite{ballard2007designing}. On account of that any kind of its imperfection is an issue. In this view usability definition provides three attributes to be considered. First, effectiveness refers to a user's ability to successfully complete a task within a given context \cite{xiong2020susapp}. Any obstacles hindering task completion are considered usability issues. Second, efficiency relates to how quickly and accurately a user can complete a task \cite{la2011efficiency}. From this perspective, any deficiencies in the application's processing capacity are considered issues as well \cite{al2021usability}.
Third, satisfaction refers to a user’s perceived level of comfort, pleasure, or fulfillment of expectations and needs \cite{palyama2022important}. From this standpoint, an issue refers to any negative emotional or subjective response a user experiences as a result of interacting with the application \cite{hajesmaeel2022most}. 

That being said, usability issues with mobile applications can affect every aspect of the application in practice. Furthermore, given the diverse and changing contexts in which users interact with applications, perceived usability falls under a much broader umbrella.
Digital distribution services such as Google Play, or the App Store have opened up new possibilities for software vendors to collect user feedback on their published mobile applications, as users can freely express their opinions on various topics, including usability \cite{iacob2013retrieving}. Indeed, many studies have analyzed online user opinions about mobile applications, providing valuable insights and enriching the existing literature on firsthand mobile usability issues. 

For example, Iacob et al. \cite{iacob2013you} found that users had difficulty configuring, using, and learning the apps.
Ismail et al. \cite{ismail2016review} discovered various issues like accessibility, accuracy, conciseness, ease-of-use, convenience, learnability, user satisfaction, task-technology fit, among many others. Weichbroth \& Baj-Rogowska \cite{weichbroth2019online} synthesized the opinions and further classified the extracted keywords into the form of seven attributes, namely: efficiency, satisfaction, effectiveness, errors, ease of use, cognitive load, and operability. Keeping in mind that over 2.08 billion users play games on their mobile devices \cite{Statista2025games}, Ho and Tu \cite{ho2011investigation} found that users paid attention to stability, ease of use, and accuracy, among other factors. 

\begin{figure}[H]
    \centering
    \includegraphics[width=9cm]{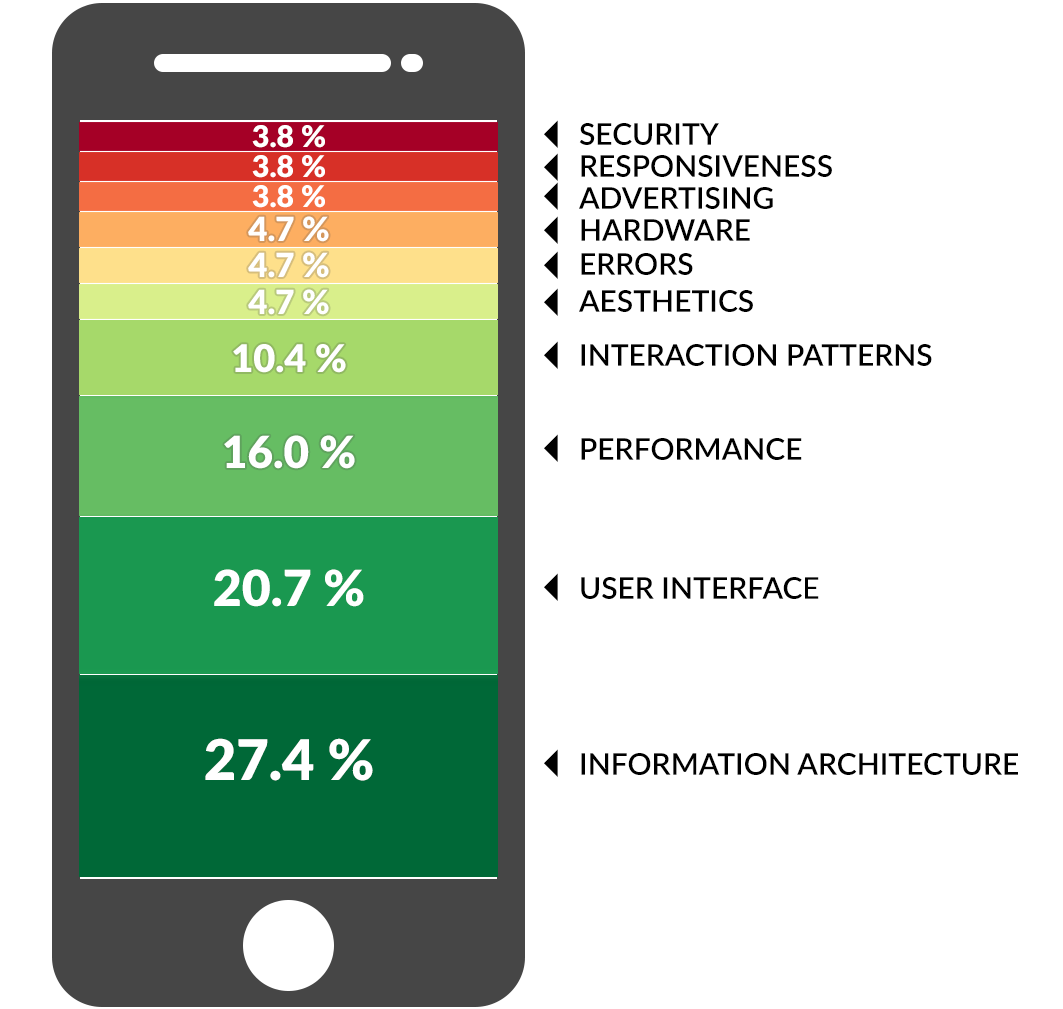}
    \caption{The ten most prevalent categories of mobile usability issues. Source: \cite{weichbroth2025usability}.}
    \label{fig:weich-resutls}
\end{figure}

In light of recent research conducted through expert interviews \cite{weichbroth2025usability}, ten categories related to the usability of mobile applications were identified, including: Information Architecture (27.36\%), User Interface (20.75\%), Performance (16.04\%), Interaction Patterns (10.38\%), Aesthetics, Errors, and Hardware (4.72\% each), as well as Advertising, Responsiveness, and Security (3.77\% each). In this regard, Figure~\ref{fig:weich-resutls} provides a graphical summary.

However, despite the maturity of usability guidelines, well-established best practices, and advanced software tools, numerous challenges and unresolved issues still need to be addressed to meet the diverse requirements of mobile users \cite{genc2017systematic}. Since, to the best of our knowledge, the current literature lacks a comprehensive perspective on this matter, we aim to identify existing countermeasures that can be applied to eliminate or mitigate usability issues in mobile application development and maintenance.

\section{Methodology}
\label{sec:methodology}
In our study we put forward the following research question: What countermeasures can be taken to avoid usability problems in mobile application development? Given the qualitative nature of the study and prior research in this area, we decided to conduct expert interviews, which allowed us to gather insights from individuals with specialized knowledge and experience.

\subsection{Identify and recruit experts}
Experts were identified and recruited by both authors using convenience sampling, a non-probability method based on accessibility and willingness to participate. This approach allowed us to efficiently engage individuals with relevant knowledge and experience in mobile application usability. In total, 21 experts were recruited.

\subsection{Interview protocol}
Since our study is organized around an explicit theme, we adopted a fully structured protocol consisting of three parts. The first part contains a brief introduction. The second part includes an open-ended question. The third part asks for demographic information.

\subsection{Interview organization}
We conducted one round of data collection, which lasted from May 2024 to February 2025. In this regard, a request with a brief description (including the research question) along with the attached file was sent to each respondent who agreed to participate in the study. Therefore, a participant could respond at any time and any place. Note that respondents did not receive any compensation.

\subsection{Data sample}
All data was collected in a single online spreadsheet using a popular and free tool from one of the world's leading software and services providers. Based on an initial assessment of the data collected, Expert \#4 was excluded from the analysis due to the cursory and generic nature of the responses submitted. The data was collected in Polish and then translated into English. Eventually, the data sample included statements provided by 20 experts, with a total of 2,021 words (11,151 characters without spaces). Table~\ref{tab:expert-sample} presents the experts' profiles along with the length of the answers provided.   

\begin{table}[]
\centering
\small
\caption{The demographic profiles of the experts and the length of their responses.}
\label{tab:expert-sample}
\begin{tabular}{|ccccccc|}
\hline
\textbf{Sex}   & \textbf{Age} & \textbf{Education} & \textbf{Current Occupation}  & \textbf{Experience} & \textbf{\#Projects} & \textbf{\#Words} \\ \hline
Man   & 39  & Higher    & Front-End   Developer        & 16        & 4        & 322   \\ 
Man   & 26  & Higher    & Product   Designer           & 5         & 2        & 200   \\ 
Man   & 42  & Higher    & UX Designer                  & 17        & 25       & 172   \\ 
Woman & 30  & Higher    & UX Designer                  & 8         & 9        & 163   \\ 
Woman & 26  & Hihger    & UX Researcher                & 4         & 3        & 160   \\ 
Man   & 25  & Higher    & Senior Data   Engineer       & 6         & 3        & 107   \\ 
Man   & 29  & Higher    & IT Team Leader               & 3         & 2        & 103   \\ 
Woman & 46  & Higher    & Mobile Front-End Dev.        & 20        & 15       & 97    \\ 
Woman & 33  & Higher    & UX Designer                  & 3         & 3        & 92    \\ 
Man   & 26  & Secondary & Front-End   Developer        & 4         & 3        & 85    \\ 
Man   & 37  & Higher    & IT Consultant                & 20        & 5        & 81    \\ 
Woman & 41  & Hihger    & UX Writer                    & 20        & 3        & 73    \\ 
Woman & 45  & Higher    & UX Designer                  & 20        & 19       & 67    \\ 
Man   & 22  & Higher    & Mobile Software Dev.         & 2         & 2        & 67    \\ 
Woman & 42  & Higher    & UX Writer                    & 16        & 2        & 62    \\ 
Woman & 26  & Hihger    & UX Designer                  & 3         & 5        & 50    \\ 
Man   & 31  & Secondary & UX Designer                  & 7         & 3        & 48    \\ 
Woman & 56  & Higher    & Software   Developer         & 20        & 15       & 45    \\ 
Woman & 41  & Higher    & Technical   Writer           & 16        & 2        & 15    \\ 
Man   & 42  & Hihger    & CPO                          & 10        & 2        & 12    \\ \hline
\end{tabular}
\end{table}

In summary, 20 experts participated in our study, including 10 men and 10 women. The average age of the men and women was 31.9 and 38.6 years, respectively. Out of 20 experts, all women and eight men have higher education, while 2 men have secondary education. In terms of professional experience, all positions held have involved mobile application development at various stages. In our view, such diversity should be seen as beneficial, as it provides a broader perspective on the research problem.
On average, women (13) had more work experience than men (9). Similarly, the average number of declared projects was higher for women (7.6) than for men (5.1).

The length of the experts' responses varies significantly, ranging from 12 to 322 words, with an average of 101.05 words per answer. Interestingly, on average, men (119.7) provided longer answers than women (82.4). However, note that its length does not always reflect its explicit information value.

\subsection{Data analysis}
After preparing the data set, we did an initial reading to get a general sense of the content. In particular, we focused on the participants' language, especially in the context of the research problem. 
The in vivo codes were extracted in two rounds.
In the first (extraction) round, we highlighted exact words or phrases used by the participants that stood out from a common language, extracting a total of 212 codes. 
In the second (pruning) round, we removed all words and phrases that were irrelevant to the research question, removing 41 codes.

\subsection{Data synthesis}
Next, the remaining 171 codes were manually stemmed, meaning that prefixes and suffixes were removed from words by converting them to their root form. Grouping and categorizing the codes was done in two rounds. 
In the first, we examined the codes, looking for similarities, patterns, or contradictions among the codes.
In the second, we grouped the codes into broader themes, while still preserving the participants' voices, and counted their frequencies. Finally, we divided the codes into two categories: guideline (an indication of what or how something should be done) and activity (things people do, especially to achieve a particular goal). Our findings include ten guidelines and three activities which are discussed in more detail in the next section.

\section{Results}
\label{sec:results}
In this section, we first present and discuss a structured framework for preventing usability issues in mobile application development. Then, we outline three activities that can be undertaken to effectively detect and eliminate these issues. 

\begin{figure}[h]
    \centering
    \includegraphics[width=13cm]{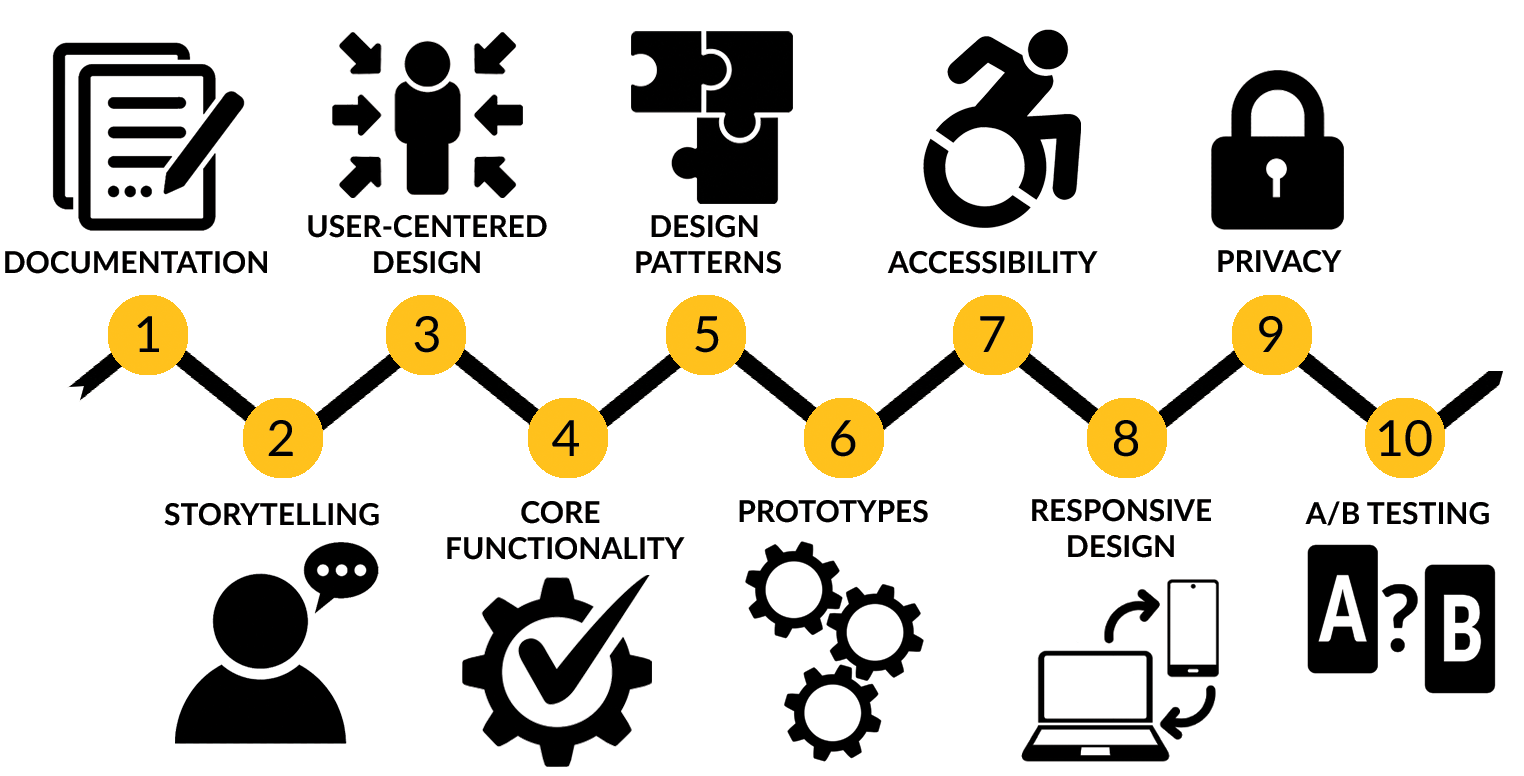}
    \caption{The MUG-10 framework for preventing usability issues in mobile application development.}
    \label{fig:10guidelines}
\end{figure}

Figure~\ref{fig:10guidelines} presents the ten mobile usability guidelines, hereafter referred to as the MUG-10 framework.

\begin{itemize}
    \item[\textbf{G1:}]\textbf{Prepare documentation}. Maintain technical, accurate, and up-to-date documentation to ensure alignment across teams, and engage stakeholders in the documentation process to ensure their perspectives are captured and reflected in product development \cite{Aggarwal2024,Mykhailov2025}. In a broader sense, well-prepared documentation supports consistent decision-making and minimizes the risk of miscommunication throughout the project lifecycle \cite{kortum2016miscommunication}.

    \item[\textbf{G2:}] \textbf{Use storytelling}. Through scenarios and personas, storytelling techniques can be used to better understand and address user needs \cite{de2023usability,Chan2024}. Storytelling builds connections, inspires action, and facilitates learning more effectively than presenting raw data alone by tapping into human emotions and shared experiences. Besides, this practice also promotes more effective communication among people by translating abstract usability narratives into specific objectives \cite{velhinho2024usability}.

    \item[\textbf{G3:}] \textbf{Apply User-Centered Design (UCD) principles}. By definition, UCD is an iterative design process that emphasizes active engagement in the mobile application development process \cite{mao2005state,trujillo2025user}. By involving users in research, testing, and continuous feedback, development teams can identify and resolve usability issues effectively. Early and frequent testing, in particular, helps ensure that the final application closely aligns with their expectations \cite{lopes2018applying}.
        
    \item[\textbf{G4:}] \textbf{Prioritize core functionality}. Mobile users typically engage with apps to quickly and efficiently perform specific tasks. Prioritizing core functionality eliminates distractions, reduces interface complexity, and improves the overall user experience \cite{Singh2024,SkyWeb2024}. Besides, by focusing development efforts on the most critical features, before adding additional functions or fine-tuning visuals, a minimum viable product (MVP) can be delivered without overwhelming users \cite{cheng2016mobile}.
    
    \item[\textbf{G5:}] \textbf{Use design patterns and templates}. Established design patterns and templates offer reliable solutions to common UI issues, reducing cognitive load by providing well-known icons, symbols, and interface structures \cite{neil2014mobile,da2022mobile}. These solutions can speed up development by giving teams reusable components that are easier to implement, maintain, and test. Using familiar patterns also improves learnability \cite{yang2024comprehensive}, as users can take advantage of their prior knowledge.
        
    \item[\textbf{G6:}] \textbf{Develop mock-ups and prototypes}. Use static, both low- and high-fidelity visuals to illustrate an application layout \cite{zhang2012comparing}. Then follow up with interactive prototypes that enable early user testing, help validate design concepts, and gather first-hand user feedback \cite{kanai2007integrated,radziszewski2021greencoin}. Graphical artifacts also facilitate better communication among designers, developers, as well as other stakeholders by providing a shared reference point \cite{wirtz2013improving}.
    
    \item[\textbf{G7:}] \textbf{Follow accessibility guidelines}. Design inclusive interfaces that follow POUR (Perceivable, Operable, Understandable, Robust) \cite{w3org-2025,Halpin2025} principles to create truly inclusive content \cite{ballantyne2018study}.
    Adhering to POUR guidelines ensures that a mobile application is usable by people with diverse abilities, including those with visual, auditory, motor, or cognitive impairments. Additionally, following these standards supports legal compliance and shields organizations from reputational harm.
    
    \item[\textbf{G8:}] \textbf{Ensure responsive design across devices}. Design and test for multiple screen sizes to provide a smooth user experience \cite{DDN2025}, while key features include flexible grids, layouts, and images that automatically adjust to different resolutions and device types \cite{almeida2017role}. Additionally, a well-implemented responsive layout reduces the need for redundant code across separate app versions, streamlining both current maintenance and future development.
    
    \item[\textbf{G9:}] \textbf{Assure user privacy}. Design and communicate transparent authorization patterns that address the permissions granted to an application to access specific data and functions on a user's device, and the means to protect the user's personal information \cite{martin2016putting}, including the right to control how such information is used, stored, and shared, while preserving the user's anonymity where possible \cite{alkindi2021user}.
    
    \item[\textbf{G10:}] \textbf{Conduct A/B testing}. Design, implement, and test two product versions to collect data-driven insights and make informed decisions based on user behavior \cite{lee2015applying,samuel2016problems}. By making decisions based on factual user feedback rather than arbitrary assumptions, the A/B testing approach helps determine the best solution \cite{Neusesser2024}.
    
\end{itemize}

In summary, our MUG-10 framework offers a comprehensive approach applicable to any mobile application development project. As can be observed, it emphasizes early planning and communication through documentation, storytelling, and mock-ups, keeping user preferences at the forefront. Prioritizing core functionality, applying User-Centered Design (UCD) principles, and using proven design patterns leads to user-friendly interfaces. On top of that, technical practices such as responsive design, accessibility compliance, and A/B testing are also key to achieving greater usability and inclusivity. Lastly, safeguarding user privacy reinforces trust, which is essential for the acceptance of technology and, ultimately, long-term user engagement.

From a practical point of view, there are three activities which can be undertaken to detect and eliminate usability issues in mobile application development. 

\begin{enumerate}
    \item \textbf{User testing}. Experts strongly emphasized involving at least five users in usability testing \cite{Glance2025} through live and moderated sessions on a regular basic, on each stage of app development \cite{samrgandi2021user}. One can also consider A/B testing with the aim comparing two versions of a design to determine which one leads to better outcomes based on actual user behavior \cite{adinata2014b}. Such approach helps reduce opinion-based changes \cite{kruger2025optimizing}, allows testing of specific variables (e.g., layout, color, wording) \cite{king2017designing}, and supports data-driven decision making to ultimately select the optimal solution \cite{goswami2015controlled}. Remarkably, low-fidelity prototypes should be used in the early stages of application design as they tend to be visually unappealing and lack detail \cite{saleh2015impact}. On the other hand, high-fidelity models should accompany feature implementation, allowing users to interact with the solution as if it were fully developed \cite{kim2020using}.

    \item \textbf{Heuristics testing}. In this approach, instead of users, experts assess an interface based on established usability principles, guidelines or rules of thumb, known as heuristics used to evaluate a user interface. Heuristics can be understood as a practical approach to quickly identify issues, but they are not fixed solutions. In this view, heuristics guide the evaluator in testing various scenarios by applying Nielsen’s 10 usability principles \cite{Nielsen1994}.

    \item \textbf{Smoke testing}. Depending on the stage of the development process and the team's workflow, smoke testing is usually performed by quality assurance testers or developers~\cite{Global-App-Testing2024}. Smoke testing is a quick way to detect critical or major defects early~\cite{herbold2022smoke}, helping avoid spending time on detailed testing when the app build is seriously flawed~\cite{reshma2022smoke}. In this sense, smoke testing can be seen as a build verification testing that precedes user testing.
\end{enumerate}

In summary, involving multiple perspectives in comprehensive usability testing throughout the design and development process makes an application usable and user-friendly. Engaging major stakeholders is not merely a recommended practice but an imperative. Ultimately, users' intention to use and adopt the mobile application in the long term undeniably confirms its quality and value.

\section{Framework Evaluation}
\label{sec:framework-eva}
Analyzing the MUG-10 framework guidelines individually raises two questions. First, how much effort is required to determine and implement a course of action? Second, what is the expected impact on user value? To answer these two questions, each guideline was independently evaluated based on available published research on mobile application usability. 

In order to structure and map the relationships between these two factors we used the impact-effort matrix tool~\cite{perez2022methodology}. Note that, in our case, effort refers to the amount of time, resources, and work required to implement the guideline, whereas impact refers to the potential positive change in the perceived usability of the mobile application by its users.

\begin{figure}[t]
    \centering
    \includegraphics[width=0.8\linewidth]{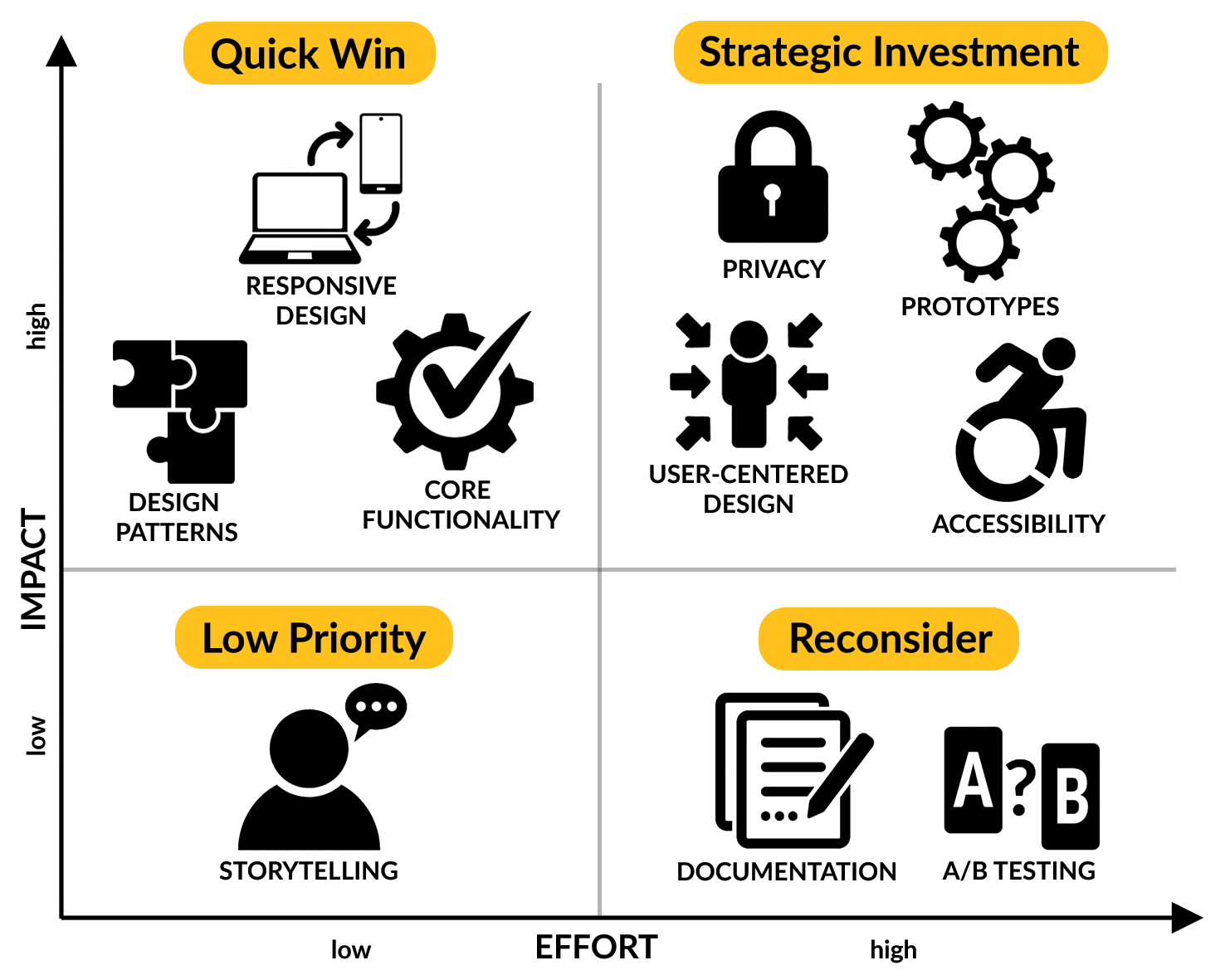}
    \caption{Impact-effort matrix of the MUG-10 framework guidelines for preventing usability issues in mobile application development.}
    \label{fig:impact-effort-matrix}
\end{figure}

Figure~\ref{fig:impact-effort-matrix} shows the detailed impact-effort matrix, which categorizes each guideline into one of four quadrants~\cite{Six-Sigma-2025}:
\begin{itemize}
    \item Quick Win: High Impact, Low Effort (upper left square).
    \item Strategic Investment: High Impact, High Effort (upper right square).
    \item Low Priority: Low Impact, Low Effort (lower left square).
    \item Reconsider: Low Impact, High Effort (lower right square).
\end{itemize}

The Quick Wins group is the highest priority and includes three guidelines that provide significant user benefits with minimal effort. First, prioritizing core functionality (G4) requires early decision-making and trade-offs but results in a user-focused list of essential app features \cite{gok2022design}, thereby reducing overall development effort \cite{teka2017user}. Second, using design patterns and templates (G5) reduces the work required for user interface design and interaction patterns by relying on tested solutions \cite{nilsson2009design,punchoojit2017usability}. Third, although ensuring responsive design across devices (G8) involves testing across various screen sizes and platforms \cite{dutson2014responsive}, it is essential for delivering a consistent user interface layout and appearance \cite{patel2015case,parlakkilicc2022evaluating}.

Second in line are Strategic Investments, guidelines that have a significant impact but require a lot of time and resources to implement. Four items fall into this category. First, applying UCD principles involves iterative testing and active user research \cite{de2014building}, leading to strong alignment with user needs and expectations \cite{sedlmayr2019user,quezada2021systematic}. Second, developing mock-ups and prototypes (G6) is time-consuming \cite{jurgensen2011motivation}, especially in the early stages \cite{jouppila2021nurses}. However, it enables early user feedback and testing \cite{duan2024generating}, which significantly reduces the risk of costly interface redesigns in the future \cite{kryszajtys2021mock}.
Third, following accessibility guidelines (G7) requires extra effort~\cite{huq2024automated}, but it ensures inclusivity and digital equality. Moreover, accessibility is a legal requirement in many regions, such as the European Union (European accessibility act), or the United States (The Americans with Disabilities Act). With over 1.3 billion people (16\% of the world’s population) living with some form of disability~\cite{WHO2023}, mobile applications that include accessible features can reach a broader user base over time.
Fourth, ensuring user privacy (G9) requires a combination of technical safeguards \cite{assal2015s}, legal compliance \cite{amaral2025gdpr}, and transparent communication \cite{asuquo2018security}. Despite requiring considerable effort, it is essential for building user trust \cite{aimeur2016changing} and meeting ethical \cite{svanaes2010usability} and legal standards \cite{hassanaly2021analysis}.

In our framework, Low Priority includes low-impact and low-effort guidelines. Only one item falls into this category: the use of storytelling (G2). It is easy to implement in presentations or personas~\cite{schleser2022mobile} and helps enhance empathy~\cite{zhu2024digital} and user understanding. However, its direct impact on usability is limited, since storytelling primarily aims to create emotional ties~\cite{yoo2015factors}, foster trust \cite{ojonugwa2022media} and loyalty~\cite{salvietti2025fostering}, and motivate users to engage more deeply with the product \cite{Thefinchdesignagency2024}.

At the bottom of the priority list are the Reconsider guidelines, which offer limited benefit to the usability while consuming substantial resources. First, preparing documentation (G1) requires considerable time and resources~\cite{redlarski2016hard}, often involving multiple stakeholders across teams~\cite{balagtas2009methodology}. However, its direct impact on usability has not yet been clearly demonstrated. Second, conducting A/B testing (G10) involves preparing two or more versions of an app~\cite{zou2024unwitting} and carrying out empirical testing with both internal and external users~\cite{lettner2013enabling}. While A/B testing is often promoted as an effective method for testing a hypothesis \cite{zarzosa2018adopting}, in many cases it is simply unnecessary, as both UI design and interaction patterns rely on proven, established solutions that may be sufficient.

\section{Discussion}
\label{sec:discussion}
\subsection{Contributions}
In light of the results discussed above, we argue that our study contributes to the current body of literature in two ways. First, our framework covers a generic and wide range of mobile usability issues, from design and functionality to testing, inclusivity, and privacy. They are well aligned with recognized best practices and, when applied systematically and according to the experience and knowledge of experts, can significantly reduce usability obstacles and burdens in mobile application development.
Second, based on a targeted literature review, we confirm the validity and usefulness of the three usability testing methods currently employed by practitioners in real-world mobile software development environments. 


\subsection{Limitations}
Nevertheless, our study, like others of a similar nature, suffers from inherent and obvious limitations. First, an interview-based study may reflect biases or assumptions rooted in the specific assumptions or requirements of the projects involved. In other words, such bias occurs when respondents rely heavily on their own experience and knowledge, which may inadvertently skew their judgments or recommendations. 
Second, our study relies on a sample size of 20 respondents, which may be considered relatively small. Basically, a small sample size may limit the reliability and generalizability of the results, meaning that the results may not accurately represent the opinions of the broader population.


\section{Conclusions}
\label{sec:conclusion}
Undeniably, usability is a key non-functional software quality that directly affects user acceptance. By contrast, poor usability can lead to user frustration and ultimately application rejection. While usability issues can arise at any stage of mobile application development, it is important to identify and understand countermeasures that can be applied effectively. To this end, we provide general guidelines and practical activities recognized by experts as working countermeasures in the conducted survey. Future research could explore how the expert-identified countermeasures perform in real-world mobile application projects through longitudinal case studies. It would also be valuable to conduct a targeted action research that evaluate the effectiveness of these recommendations, validating them across different stakeholder groups. In addition, future studies could focus on developing automated tools or frameworks to support the early detection and mitigation of usability issues during the mobile application development.


\end{document}